# Flexible and Extensible Digital Object and Repository Architecture (FEDORA)

Sandra Payette and Carl Lagoze

Department of Computer Science
Cornell University
{payette,lagoze}@cs.cornell.edu

**Abstract.** We describe a digital object and respository architecture for storing and disseminating digital library content. The key features of the architecture are: (1) support for heterogeneous data types; (2) accommodation of new types as they emerge; (3) aggregation of mixed, possibly distributed, data into complex objects; (4) the ability to specify multiple content disseminations of these objects; and (5) the ability to associate rights management schemes with these disseminations. This architecture is being implemented in the context of a broader research project to develop next-generation service modules for a layered digital library architecture.

## 1.0 Introduction:

A fundamental requirement of an open architecture for digital libraries is a reliable and secure means to store and access digital content. FEDORA is a digital object and repository architecture designed to achieve these requirements, while at the same time providing extensibility and interoperability. The key features of the architecture are: (1) support for heterogeneous data types; (2) accommodation of new types as they emerge; (3) aggregation of mixed, possibly distributed, data into complex objects; (4) the ability to specify multiple content disseminations of these objects; and (5) the ability to associate rights management schemes with these disseminations.

FEDORA is positioned within a larger open-architecture framework in which the total functionality of a digital library is partitioned into a set of services with well-defined interfaces. These core services include: (1) *repository services* that provide the mechanisms for depositing, storing and accessing digital objects; (2) *index services* that provide the mechanisms for discovering digital objects; (3) *collection services* that provide the means of aggregating sets of digital objects and services into meaningful collections; (4) *naming services* that register and resolve globally unique, persistent names for digital objects; and (5) *user interface services* that provide a human gateway into the other services. The well-defined interfaces of these core services allow them to be combined with each other and other value-added services to create usable instantiations of digital libraries.

This multi-layered service structure evolves from the concepts implemented in the Dienst Architecture [9], which is the foundation for the Networked Computer Science Technical Report Library (NCSTRL) [12]. Currently, the Digital Library Research Group at Cornell University is engaged in a number of research projects to develop next-generation service modules for a layered digital library architecture. FEDORA addresses the requirements for digital objects and the repository service that provides access to them.

In this paper we describe the FEDORA architecture. Section 2 provides an overview and introduces the three logical layers of the architecture. The next three sections describe the architecture of a Digital Object: Section 3 describes the structural layer of the architecture; Section 4 describes the interface layer; and Section 5 focuses on rights management for Digital Objects. Section 6 describes the Repository, which represents the management layer of the architecture. In Section 7, we provide a practical scenario of how the architecture works, and in Section 8 we conclude and describe our future work.

## 2.0 FEDORA: An Overview

The FEDORA architecture describes a content container abstraction, the *DigitalObject*, and the nature of *Repositories* that provide access to these containers. We conceptualize a *DigitalObject* as having: (1) a structural kernel, which encapsulates content as opaque byte stream packages and, (2) an interface, or behavior, layer that gives contextual meaning to the data in the DigitalObject. One useful metaphor for a DigitalObject is that of a cell. At the core is a nucleus containing essential data. Surrounding this structural nucleus is a functional layer containing content *Disseminators,* components that transform core data packages into recognizable information entities such as books, multimedia encyclopedias, and the like. Both the kernel/nucleus and the content disseminators are wrapped within the cell membrane that marks its contents as a uniquely identifiable DigitalObject.

For example, a simple DigitalObject might have a structural kernel that contains a number of byte stream packages that are `gif` images and another byte stream containing Dublin Core [4] metadata. On top of this structural layer there might be an interface layer that endows the DigitalObject with book-like behavior, allowing a client to access the table of contents or a specific page. The same DigitalObject might also have descriptive metadata behavior, allowing access to bibliographic fields such as the book's author or title.

The *Repository* is the FEDORA component that provides for the management of and access to named DigitalObjects. From the perspective of the Repository, the DigitalObject is a generic opaque entity known only by its unique name. The Repository provides services to deposit, store, replicate and access DigitalObjects as generic components.

### 2.1 Extensibility for Types and Rights Management

The architectural layers discussed above are key to the FEDORA extensibility model. By segregating structure from interfaces, FEDORA makes it possible for DigitalObjects of extreme structural variation to present themselves to clients in a "normalized" manner. From a client perspective, the interface layer provides the means to interact with DigitalObjects through a set of behaviors that define global or domain-specific notions of "content." A set of behaviors that formally specifies the nature of a particular form of content is referred to in FEDORA as a *content type*. Content types provide the mechanisms for establishing functional equivalence among DigitalObjects with disparate internal structures. In addition, they provide a simple client-oriented interface to the object that hides the structural complexity contained within.

Extensibility of content types is of paramount importance in the repository architecture. There are already countless forms of content and new ones will continue to appear. Any viable architecture must seamlessly integrate new content forms, and the mechanisms for disseminating and presenting them. To promote extensibility, FEDORA does not predetermine any taxonomy of content types. Instead, FEDORA creates the means to link external types to DigitalObjects. The FEDORA content-type system is extensible because content types are, themselves, named entities in the digital library infrastructure that can be referenced by a unique identifier such as a Uniform Resource Name (URN).

This same extensibility model applies to rights management schemes. FEDORA's facilities for access management are inclusive rather than prescriptive - accommodating a variety of existing and new rights management schemes, rather than attempting to promulgate one global scheme.

### 2.3 Theoretical Foundations and Related Work

The theoretical foundations for FEDORA lie in the Kahn/Wilensky framework [7] and the Warwick Framework [8], both of which influenced the design of the FEDORA DigitalObject. FEDORA's extensibility model originates in the Distributed Active Relationship (DAR) abstraction, described by Daniel, Lagoze, and Payette[5]. DARs leverage the connectivity and computational characteristics of networked environments to create dynamic relationships between data resources. These relationships: (1) can exist between resources in different repositories, (2) can be executable, and (3) can be named entities themselves in the infrastructure.

Others are working to develop architectures and document models that support extensible document types. Monch, Drobnik and Wolfgang [11] propose a document architecture that supports the integration of new media types and formats into digital library documents. The U.C. Berkley DLI project has developed the "multivalent document model" [13] in which documents are viewed as layers of content supported by dynamically loaded behaviors. FEDORA, in contrast to these projects, is distinguished by its architectural segregation of DigitalObject structure, content-type interfaces, and mechanisms that execute content-type behavior and by its attention to rights management. FEDORA also provides the mechanisms to identify, store, and access these types in the infrastructure.

Parts of the FEDORA work are the result of a collaboration with the Corporation for National Research Initiatives (CNRI) which is implementing a Digital Object Repository for the Library of Congress Digital Library Initiative based on their Repository Access Protocol (RAP) [1]. Another related effort is the Making of America Project (MOAII) [10], which has proposed a Digital Library Service Model to support a distributed digital library of archival materials. This project, coordinated by the Digital Library Federation (DLF), is developing standards for creating and encoding digital archival objects such as books, diaries, and photographs, and is specifying a service model for each. We are collaborating with members of the DLF to demonstrate the extensibility of FEDORA by accommodating these community-specific types as they emerge.

### 2.4 Implementing the Architecture

FEDORA is being implemented in CORBA and Java, however we believe the abstractions and service requests we have developed to be of general applicability and to be able to transcend particular implementation technologies. We are initially using a subset of the NCSTRL collection as our testbed. By working closely with CNRI to achieve architectural convergence between the RAP and FEDORA projects, we hope to demonstrate interoperability across independently-developed repository architectures that use different underlying CORBA Object Request Brokers (ORBs).

## 3.0 The Structural Layer (Digital Object Structural Kernel)

The lowest layer of the Repository architecture supports examination and manipulation of the structural composition of a *DigitalObject*. At this layer we introduce abstractions and service requests to compose, manipulate and access DigitalObjects in a generic manner. The DigitalObject is an abstraction for aggregating heterogeneous data, stored as typed byte stream packages known as *DataStreams.* Structural access to this set of DataStreams is provided by a set of basic service requests, collectively known as the *Primitive Disseminator.* These service requests provide a means to operate on diverse and heterogeneous data in a consistent manner. The following architectural abstractions make up the kernel of the DigitalObject:

**DataStreams**

A DataStream is a typed byte stream that preserves the internal format and encoding of the type, but encapulates it so that it can be treated generically within the DigitalObject. This distillation of data into its essential type and byte representation allows heterogeneous forms of digital content to be treated in a uniform manner, essentially creating a state of interoperability at the level of the DigitalObject, instead of at the level of the individual content packages themselves.

Any type of data can be represented as a DataStream, and disparate forms of data can be aggregated in a DigitalObject as DataStreams. For example, the same DigitalObject can contain byte stream representations of a TIFF image of a photograph of a rare bird, a Dublin Core record describing the image, a digital audio track of the bird's song, and a geographic dataset that shows the nesting locations of the bird. At this structural layer, the contextual relationships among DataStreams are not defined. Thus, DataStreams can not be distinguished by their semantic roles such as data or metadata. These contextual relationships are defined at the interface layer, which is described in Section 4.

**Primitive Disseminator**

From a client perspective the DigitalObject is a sealed wrapper, known only by it unique name (e.g., a URN), that can be manipulated through a set of defined service requests. The result of any of these service requests is a *dissemination*, a view of the information contained within the DigitalObject. From the rights management perspective, which we will discuss later, the ability to obtain a dissemination can be managed via access control mechanisms.

A *Disseminator* is an abstraction for packaging a set of service requests that release disseminations from a DigitalObject. These disseminations are byte streams that may be content (e.g., a page of a book), an applet (e.g., a viewer for some content), or a mixture of both. The *PrimitiveDisseminator* is the set of service requests that is common to all DigitalObjects. (We will describe other content-specific Disseminators in Section 4). Within the kernel, the PrimitiveDisseminator provides the fundamental set of requests that allow access to the structural layer of the FEDORA architecture. These requests fall into three categories: (1) those that support the composition and manipulation of DigitalObjects, (2) those that support the access to the DigitalObject structure and its internal DataStreams, and (3) those that serve as a common mechanism for adding, discovering and invoking content-specific behavior in the context of the DigitalObject. The requests in this third category can be thought of as providing a "gateway" to the interface layer of the architecture. Table 1 lists the major service requests from the above categories, and describes their functionality.

From an object-oriented standpoint, the PrimitiveDisseminator is really the set of methods defined for the class DigitalObject. The CORBA Interface Definition Language (IDL) description of the DigitalObject can be found in [6].

**Table 1.** Service Requests in the PrimitiveDisseminator for a DigitalObject

| **For composition:** | |
|---|---|
| CreateDataStream | Takes a file or a stream and transforms content into a typed DataStream which is inserted into the DigitalObject kernel. |
| SetAccessManager | Associates an external rights management scheme with |
| | the PrimitiveDisseminator by referencing an object that contains an executable. (See Section 5.0) |
| **For structural access:** | |
| GetDataStreams | Returns a sequence of references to DataStreams within the DigitalObject. This request exposes the number and types of contained DataStreams. The actual stream of bytes within each DataStream can be accessed, however, the request will not provide information on the meaning or the context of the data. |
| GetAccessManager | Returns a reference to the AccessManager component to provide manipulation of its attributes. (See Section 5.0) |
| Gateway to content layer: | |
| CreateDisseminator | Creates a content-specific disseminator for the DigitalObject by associating one or more DataStreams with a particular set of content-specific behaviors or service requests. A set of service requests that pertains to a particular content type is a named entity that can be referred to by a unique identifier (e.g., a URN). |
| GetDisseminators | Returns a sequence of references to Disseminators associated with the Digital Object. Once a reference to a Disseminator is obtained, it can be used to manipulate and access its internal attributes. |
| ListDisseminatorTypes | Returns a list of disseminator types, or content types, associated with a Digital Object. Content types are represented by a unique identifier such as a URN. |
| ListDisseminatorMethods | Takes a disseminator type as input and returns a list of service requests (methods) that are specific to the content type. |
| GetDissemination | Takes a dissemination service request as input and returns the stream of bytes produced by the invocation of that request. Essentially, returns a content-specific manifestation of DataStreams contained in the Digital Object. |

Figure 1 depicts the core structure of a FEDORA DigitalObject at the kernel layer. It aggregates three byte stream data packages: a Postscript stream, a MARC record, and an Access Control List (ACL). These packages are depicted in the diagram as MIME-typed ByteStreams with no identifiable relationships among them. The PrimitiveDisseminator also resides in the structural kernel to endow the DigitalObject with its fundamental set of behaviors. As shown, the PrimitiveDisseminator is the single window into the DigitalObject kernel.

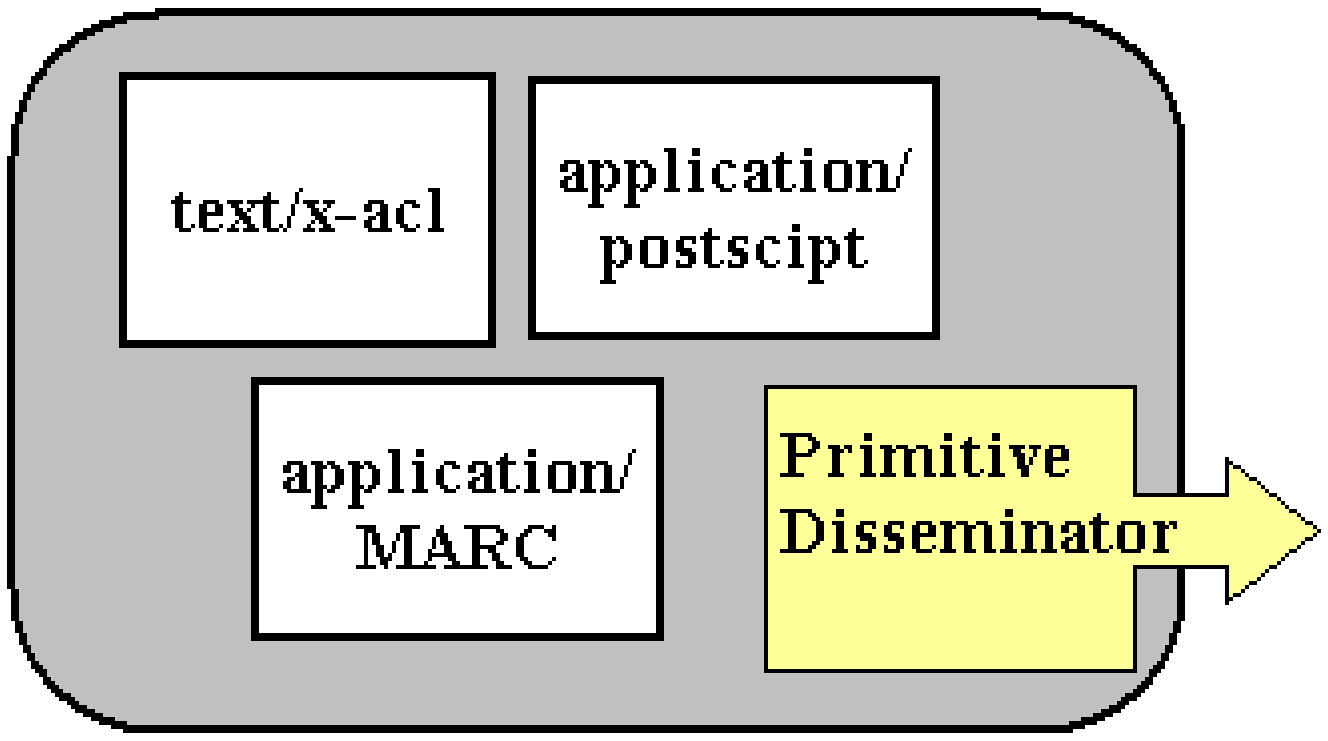

**Fig. 1.** Digital Object Structural Kernel

## 4.0 The Interface Layer (Digital Object Content-Type Disseminators)

The PrimitiveDisseminator provides service requests to interact with the DigitalObject as a structural entity. Clients, in most cases, should be insulated from the structural aspects of a DigitalObject. Instead they should interact with DigitalObjects as manifestations of well-known content forms such as a books, journals, or movies. Correspondingly, content creators should be able to restrict access to the raw byte streams of a DigitalObject by endowing it with "allowable" behaviors that represent a particular form of content (known globally or within a particular domain).

We define a *content type* as a set of service requests that specify the behaviors of a particular form of content. For example, a "book" is a content type that has operations such as `getTableofContents`, `nextPage`, or `nextChapter`. Similarly operations such as `nextArticle` and `nextIssue` are characteristic of a "journal".

A single DigitalObject can have multiple content types. For instance, the same DigitalObject may have a Dublin Core content type as well as a journal-article content type. In this case, a client could discover both content types and invoke the service request of either, without knowing anything about the DigitalObject's internal structure and configuration.

In fact, to reinforce the distinction between the internal structure of a DigitalObject and the content manifestations it can produce, we introduce the notion of *content-type equivalence*. Two DigitalObjects with entirely different structural compositions can have the same content types associated with them. A simple example of this can be demonstrated with the Dublin Core content type, formally represented by the service requests `getDCField` and `getDCRecord`. One DigitalObject may contain an actual Dublin Core record stored in a DataStream. Another DigitalObject may contain a MARC record and a mechanism to transform MARC into Dublin Core. Both objects could have Dublin Core content types associated with them, enabling each to produce the same output in response to the formal Dublin Core service requests. The *structure* of DigitalObjects, and the *mechanisms* for disseminating the content from structure, are opaque to a client, which only interacts with the DigitalObject through its *content types*.

Content-type functionality is enabled at the *interface layer* of the architecture by *content-type Disseminators*. Each content-type Disseminator endows a DigitalObjects with a set of extended service requests that pertain to a content type. As such, these components provide service requests to disseminate custom renderings of the data contained in the DigitalObject, without exposing its underlying structure.

**The Content-Type Disseminator**

In addition to the Primitive Disseminator, which is logically associated with every Digital Object, a content creator may choose to associate one or more content-type Disseminators with a DigitalObject. Each Disseminator that is associated with a Digital Object identifies a particular content type, and a set of DataStreams to be used as arguments when executing the service requests that define the type. Figure 2 depicts a DigitalObject with three Disseminators attached to it, each linking to the Datastream(s) required as input for its operations. Essentially, the DigitalObject in the figure can be treated as a book, a MARC record, or Dublin Core metadata.

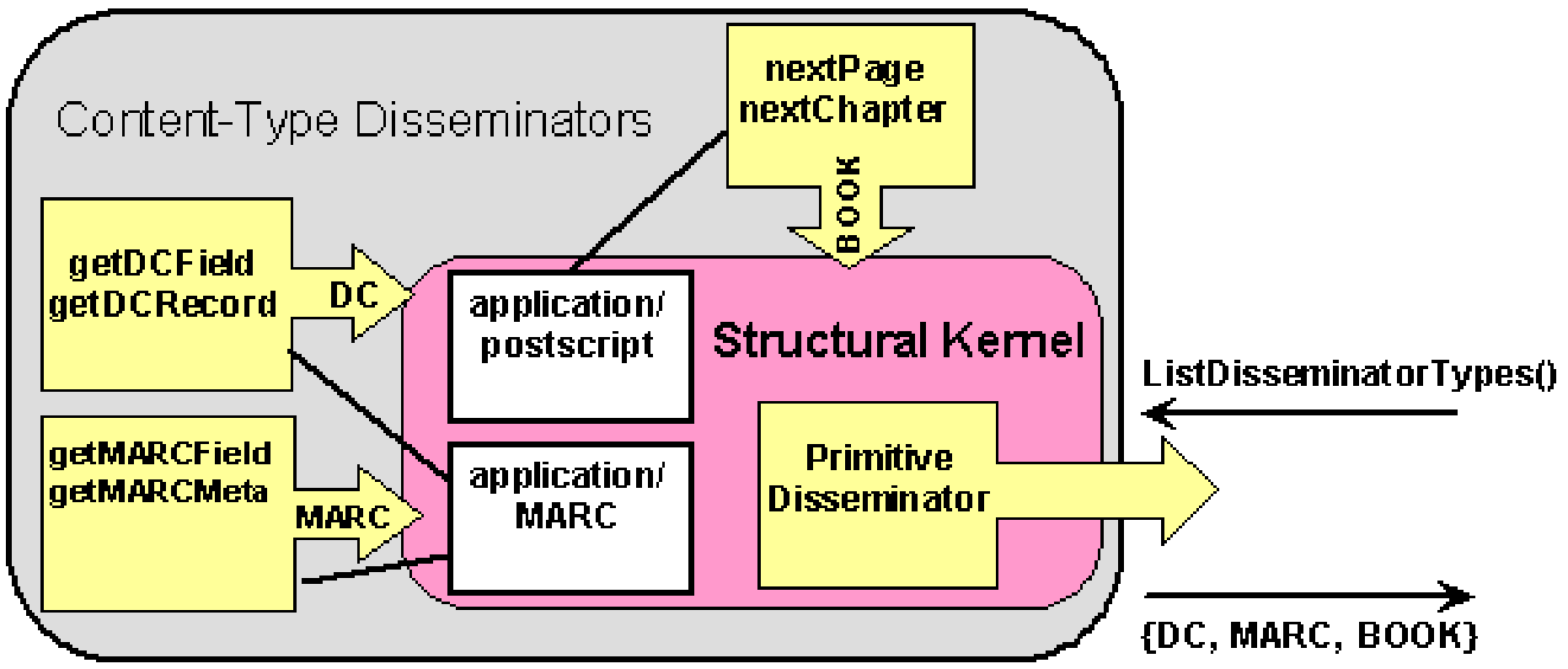

**Fig. 2.** A DigitalObject with three Disseminators

Clients do not speak directly to a Disseminator, instead they obtain information about content types, and initiate content-type operations, by invoking the "gateway " operations of a DigitalObject's PrimitiveDisseminator. For example, by issuing the `ListDisseminatorTypes()` request, a client can obtain a list of all content types associated with the DigitalObject (see lower right corner of Figure 2). Clients also use the PrimitiveDisseminator to invoke the service requests associated with these content types. Specifically, clients issue the `GetDissemination` request, with a content-type service request as an argument, to invoke a desired content-type behavior on a DigitalObject. This essentially creates an encapsulated service request, for example: `GetDissemination(BOOK.nextPage()).`

The service requests of particular content-types are *not* actually specified in the DigitalObject architecture; instead, FEDORA provides the *means to link to* externally-defined content types. This is intentional and a necessity for an extensible type system. The next section describes *how* content-specific types become named entities in the digital library infrastructure.

### 4.1 The Extensibility Model for Content Types

In the previous section we described an abstraction, the Disseminator, for associating content-type behaviors with DigitalObjects. The PrimitiveDisseminator provides the service requests for discovering the content-types associated with an individual DigitalObject. In addition, we described how a single content type (for example, Dublin Core) can be manifested in a number of structurally different DigitalObjects using different mechanisms to translate the underlying DataStreams to the disseminations of the type.

The use of external content types, however, essentially creates a "registry problem." How are these types and mechanisms uniquely named so clients can recognize and access them?

The answer to the registry problem lies in the facilities already provided by FEDORA, in conjunction with a global name service such as CNRI's Handle System [2]. Both content types and their mechanisms are, themselves, forms of content. A content type is expressed as a *signature*, the set of service requests particular to that type. A mechanism is a program or agent for executing those service requests on a set of arguments. Abstractly, both signatures and mechanisms are forms of content. As such, they can be stored and disseminated in the same fashion as any form of content, by a DigitalObject. Since a DigitalObject has a unique name (e.g., a URN) - and these names can be registered with a global naming service - content types become identifiable by the unique identifiers of the DigitalObjects that disseminate them. Thus, the registry problem is solved by using uniquely-named DigitalObjects as the means of storing and disseminating content-type signatures and mechanisms.

This section introduces two new abstractions, the *SignatureDisseminator* and the *ServletDisseminator*, to support the dissemination of content types and their mechanisms.

**SignatureDisseminator**

Each content type is specified as a set of service requests that operationally define it. The formal description of a set of content-type service requests is referred to as a *ContentTypeSignature*. The signature defines the name and syntax of each service request for the content type; for example, `nextPage()` and `nextChapter()` might be the simple signature for the type `BOOK`. To promote availability and extensibility of content types, a ContentTypeSignature is available as a dissemination of a DigitalObject, and its identity is the unique name of that DigitalObject.

A DigitalObject that contains a content-type behavior specification will have a special class of Disseminator attached to it: the *SignatureDisseminator*. This special

Disseminator is only used with DigitalObjects that release a content-type specification. It is the architectural component that enables communities to develop their own definitions of content types and make them available for general use (in conjunction with the ServletDisseminator discussed below). FEDORA defines the special SignatureDisseminator to endow a DigitalObject with the `getSignature()` request that returns a set of method specifications for a content type.

**ServletDisseminator**

The *ContentTypeServlet* is a mechanism that executes the set of content-type service requests defined by a particular ContentTypeSignature, using a specific set of DataStreams as arguments. Like the ContentTypeSignature, a ContentTypeServlet is available as a dissemination of a DigitalObject, and is identifiable by the unique name of the object that releases it.

Figure 3 shows how the ContentTypeSignature and ContentTypeServlet are used to provide content-type behavior to a DigitalObject.

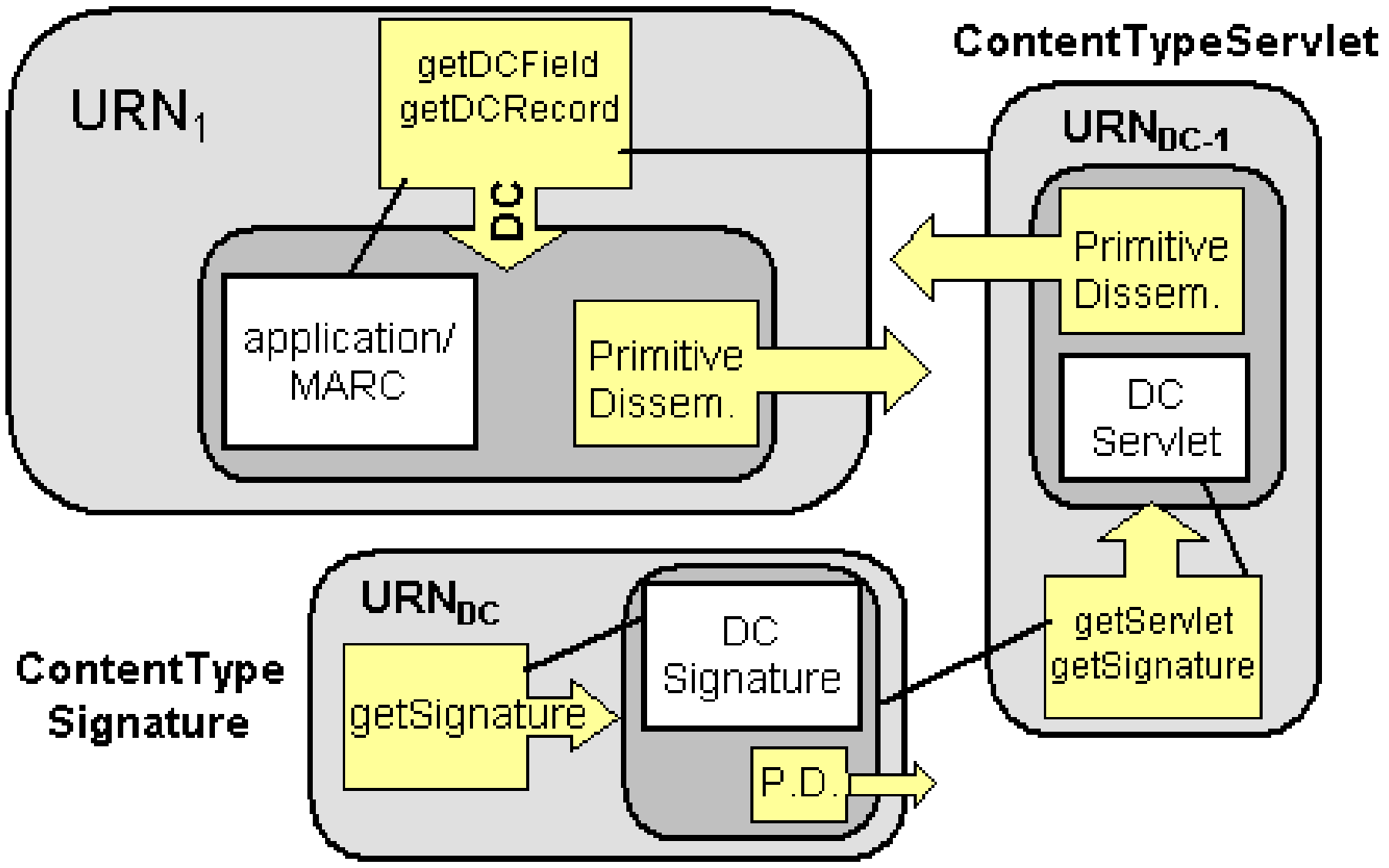

**Fig. 3.** Disseminator references a Servlet and Signature

In the figure, a Disseminator (labeled DC for Dublin Core) is created in the DigitalObject named $URN_1$ to endow it with Dublin Core content behavior. This is done by associating the DataStream containing the MARC record with a Disseminator that references the Dublin Core content type. In the example, the Disseminator directly references a servlet (mechanism) that converts MARC records to Dublin Core. This servlet is disseminated from the DigitalObject labeled $URN_{DC-1}$. Note that the ServletDisseminator of $URN_{DC-1}$ references another DigitalObject named $URN_{DC}$ that contains the formal ContentTypeSignature for Dublin Core. The net result is that a GetDissemination request on $URN_1$ activates the servlet to process the MARC record in the DigitalObject. From the client's perspective, the content type available from $URN_1$ is $URN_{DC}$ and the fact that this is enabled by the mechanism $URN_{DC-1}$ is opaque.

It should be noted that there can be a one-to-many relationship between ContentTypeSignatures and ContentTypeServlets. A single signature can be implemented by multiple servlets using different underlying techniques and data structures. For example, consider the content type BOOK. The BOOK signature may be implemented by two different ContentTypeServlets. The first servlet may fulfill the nextPage() request by performing a TIFF to GIF conversion on the DataStream that contains the TIFF image for the page. The second servlet may simply return a DataStream that contains the static GIF representation of the page. In either case, the servlet could fulfill a specified requirement that nextPage() must return a single page in the GIF format.

The CORBA IDL for each of these new abstractions - the Disseminator, SignatureDisseminator, and ServletDisseminator - can be found in [6].

### 4.2 Creating a Content-Type Disseminator for a DigitalObject

The book example, above, demonstrates that different implementations of the same content type may use different procedures and data formats to achieve equivalent results. To effectively incorporate a particular content-type (such as BOOK) in a DigitalObject, creators of DigitalObjects must have an easy way to assess the *data requirements* of the particular servlet that implements that content type. Accordingly, every ContentTypeServlet has an *AttachmentSpecification* to describe the kinds of DataStreams that must be present for the servlet to successfully execute.

Each ContentTypeServlet can define its own underlying data structure within the framework of the AttachmentSpecification. An AttachmentSpecification is an ordered sequence of *AttachmentStructures*. Each AttachmentStructure specifies: (1) structure identifier, which could be used to describe a role that a DataStream fulfills, (2) a MIME type for the DataStream(s), and (3) an ordinality indicator that specifies the required number of each type of DataStream.

Figure 4 shows a very simple AttachmentSpecification. In the example, the AttachmentSpecification is essentially a template for a hypothetical content type called PhotoAlbum. This type aggregates a set of images and thumbnails and allows browsing of them. Each column of the template contains instances of the data elements of the AttachmentStructure (a structure id, a MIME type, and an ordinality indicator).

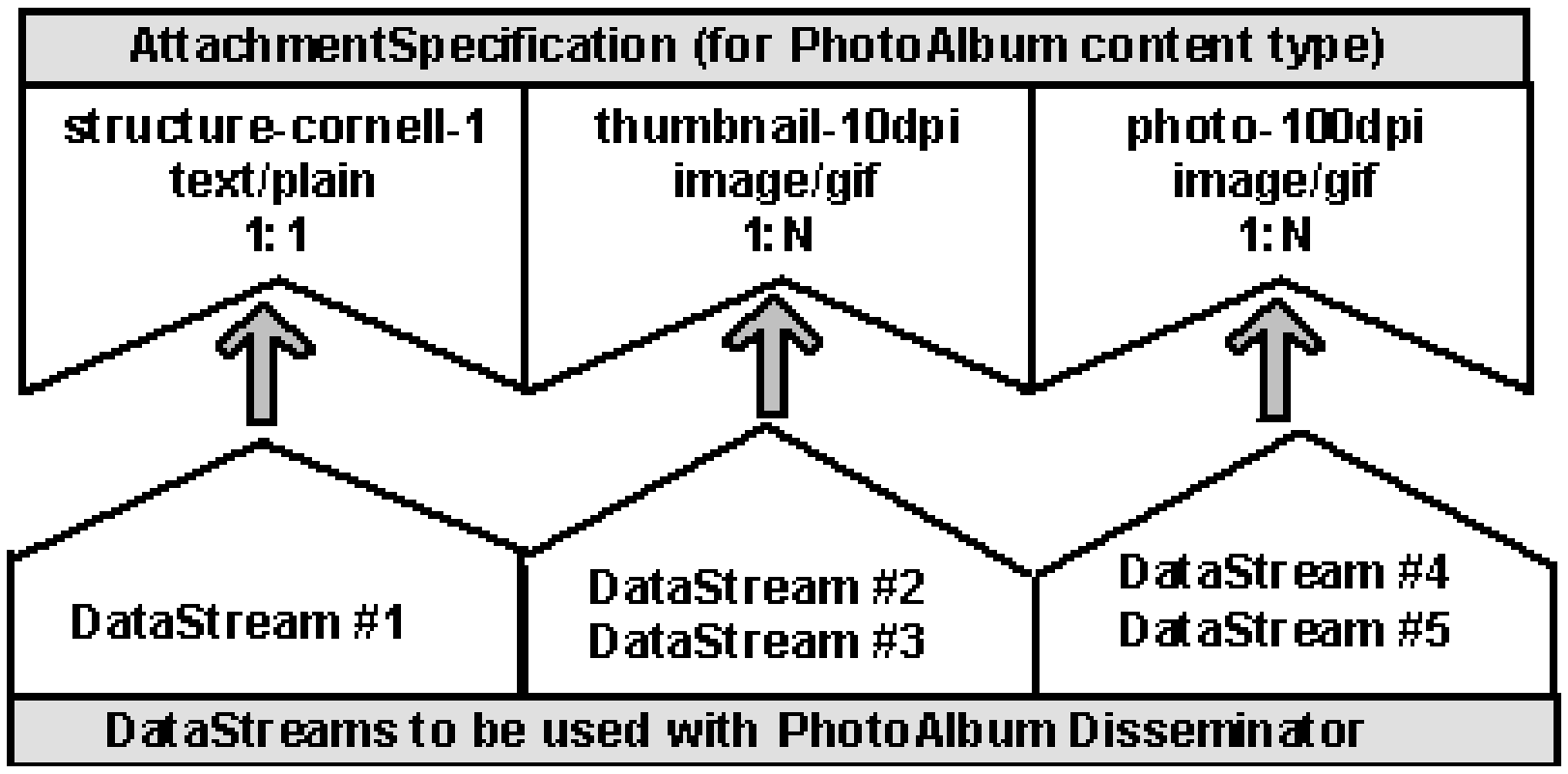

**Fig. 4.** Associating DataStreams with a Disseminator using the AttachmentSpecification

The above AttachmentSpecification tells a DigitalObject creator that the *first* attachment must be a single (1:1) DataStream containing a structure file that correlates thumbnail images of photographs with their respective exhibit-size images. In the example, the structure file is a domain-specific format known as `structure-cornell-1`, however, it could be a widely accepted structural metadata format. The *second* attachment must be a sequence of one or more (1:N) DataStreams of MIME type `image/gif` that will play the role of thumbnails for the photographs. The *third* attachment must be a sequence of one or more (1:N) DataStreams of MIME type `image/gif` that are the actual photographs. In this example, the author will use the structure file to correlate thumbnails to full images using the internal identifiers of the DataStreams.

This basic approach can be applied to create more complex Disseminators that operate on multiple sets of inter-related DataStreams of varying MIME types. The basic Disseminator architecture remains simple; it accommodates externally developed definitions of content types and external mechanisms for executing content-type behaviors.

We plan to develop a GUI tool that will easily allow object creators to associate Disseminators with DigitalObjects based on these AttachmentSpecifications. The CORBA IDL for the AttachmentSpecification can be found in [6].

## 5.0 Rights Management for DigitalObjects

We have thus far described an architecture for encapsulating data and defining content-specific interfaces for DigitalObjects. It is essential that the architecture provide facilities to protect the intellectual property that may be encapsulated in these DigitalObjects.

In FEDORA, AccessManagers are the entities that enable the association of rights management mechanisms with service requests that can be made on a DigitalObject.

Extensibility of rights management schemes is achieved in a manner similar to that already described for content types.

**The AccessManager**

The AccessManager is a generic component for attaching a rights management mechanism with a Disseminator linked with a DigitalObject. Each Disseminator, including the PrimitiveDisseminator, can have an AccessManager associated with it. The AccessManager associated with a Disseminator may internally provide different levels of access control for each service request defined by that Disseminator.

When the PrimitiveDisseminator has an AccessManager associated with it, structural-level service requests are processed under its control. This AccessManager may have different levels of access control for the individual operations defined by the Primitive Disseminator. For example, the service requests that allow creation and manipulation of content in the DigitalObject kernel may not be available to a general client; these will typically be reserved for those who author and manage DigitalObjects. Access to pure byte streams (DataStreams) can be provided to some or all clients; however, we believe that most clients will be interested in accessing DigitalObjects at the interface level through their content-type Disseminators. This implies that the "gateway" service requests of the Primitive Disseminator, such as `ListDisseminatorTypes` and `GetDissemination`, will be available to most access-oriented clients.

The ability to associate AccessManagers with content-type Disseminators provides an additional level of rights management. When a client invokes a service request associated with a content-type Disseminator the AccessManager is activated. The effect is that the service request is processed under the full control of the Access Manager. This control can be both at the point of invocation of a service request and at the point of transmission of the results of the service request back to the client.

For instance, using our previous Book example, an AccessManager may enforce payment of five cents per page (by interacting with some digital payment system) before invoking the `getPage` service request. In addition, it may apply a digital watermark to the disseminated page before it is transmitted back to the client.

## 5.1 Extensibility for Rights Management

To provide extensibility for rights management mechanisms, FEDORA uses the same approach as it does for content-type behaviors. Essentially, an AccessManager looks much like a Disseminator in that it has a type and a set of associated DataStreams. As with a Disseminator, these DataStreams are arguments to the execution of the AccessManager. For example, an AccessManager that provides access-control-list functionality would require a DataStream containing the ACL specific to the DigitalObject whose content we want to protect.

Like a Disseminator, an AccessManager identifies a type, the *AccessManagerType.* This type is the unique name (URN) of the DigitalObject that disseminates the mechanism to execute a particular rights management scheme. This mechanism is known as an *AccessManagerServlet.* Storing rights management mechanisms in DigitalObjects allows the use of external or third-party mechanisms for rights management. It also accommodates future extensibility allowing new rights management schemes to be created and stored in the infrastructure.

Figure 5 depicts a DigitalObject named $URN_1$ that has a Dublin Core (DC) content Disseminator guarded by an AccessManager. (AccessManagers are shown as 3-D boxes surrounding Disseminators.) The AccessManager is of type $URN_{ACL-1}$, which is an access control list scheme. This type is actually the unique name of the DigitalObject that disseminates the mechanism, the AccessManagerServlet, for the ACL scheme.

In the figure we can see that the AccessManager obtains the AccessManagerServlet from a DigitalObject named $URN_{ACL-1}$. This DigitalObject uses a special Disseminator, called the *AccessManagerDisseminator*, to make the ACL servlet available. Also, note that the AccessManager links to the DataStream named $DS_1$. This DataStream contains an access control list for DigitalObject $URN_1$ that is used as input to the ACL servlet. The effect is that a client invocation of either of the service requests defined for the Dublin Core content Disseminator will occur under the control of the access- control-list rights management scheme, using the ACL data contained within the $DS_1$ DataStream.

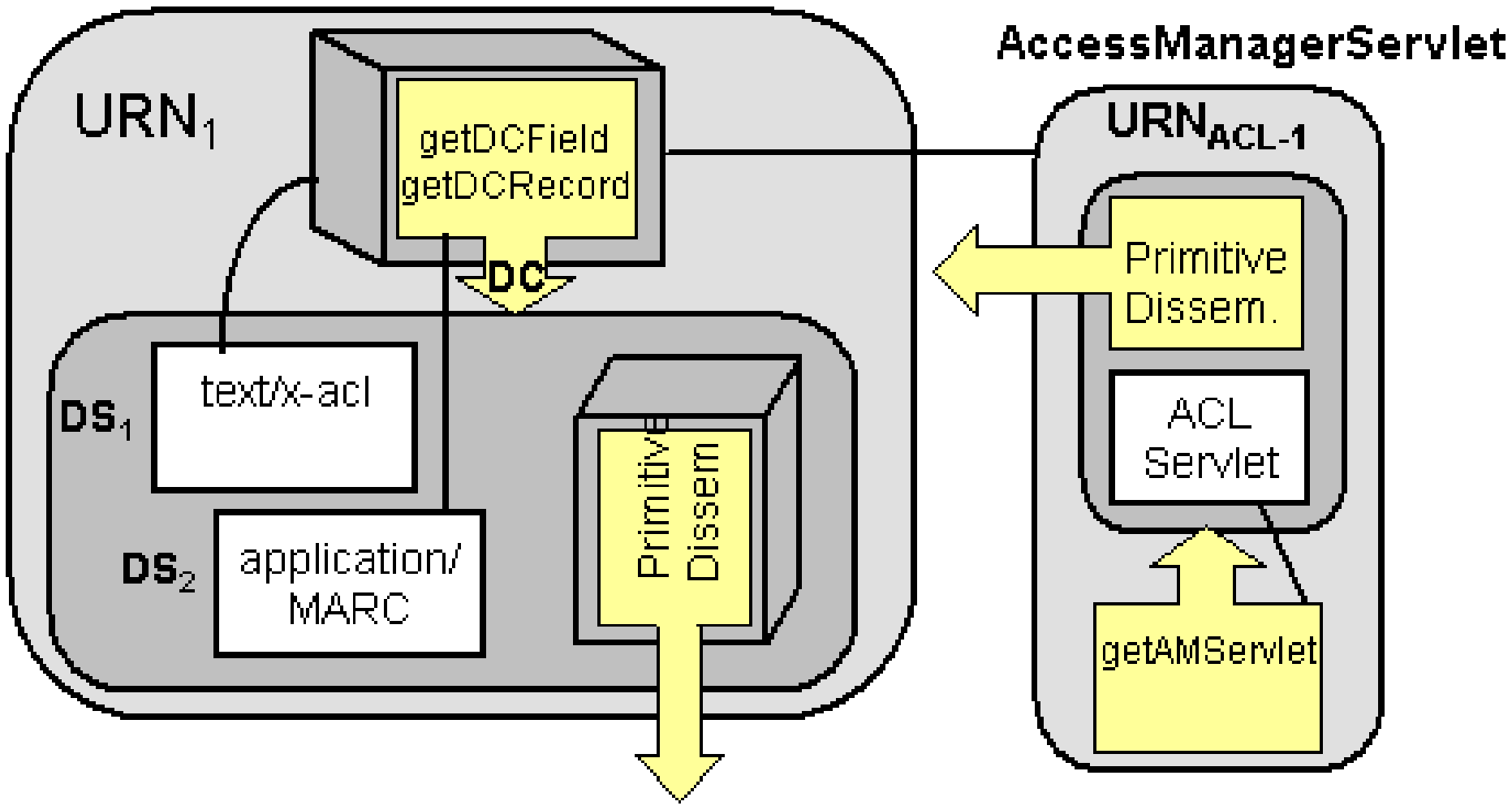

**Fig. 5.** Using an AccessManager with a Disseminator

## 6.0 The Management Layer (*DigitalObject Repository*)

Up to this point we have discussed the architecture of the DigitalObject that specifies components and service requests to manipulate and access the contents of an individual object. To provide a service context in which these DigitalObjects live, there must be an architectural layer responsible for managing DigitalObjects.

DigitalObjects cannot spontaneously create and destroy themselves; they cannot independently move themselves around in the information infrastructure; nor can they execute their content specific behaviors independently. The third layer of the FEDORA architecture provides these *object lifecycle* and *management* functions. The fundamental abstraction at this layer is the Repository itself.

**The Repository**

The *Repository* is the entity that provides management of and access to *contained* DigitalObjects. It also provides the environment in which ContentTypeServlets and AccessManagerServlets are executed. It is not a physical entity in which DigitalObjects actually reside. Instead, it is a service layer that presides over a logical grouping of DigitalObjects.

At the Repository level, DigitalObjects are completely opaque, meaning that there are no service requests that allow "looking into" the DigitalObject wrapper. Operations performed at this layer include moving a digital object, replicating it, deleting it, and others as described in [3]. As indicated in Figure 6, the only attribute of the DigitalObject that is visible at the Repository layer is its unique identifier, which is registered with a naming service of the digital library infrastructure. This figure also illustrates the "move" operation of a DigitalObject from one repository to another.

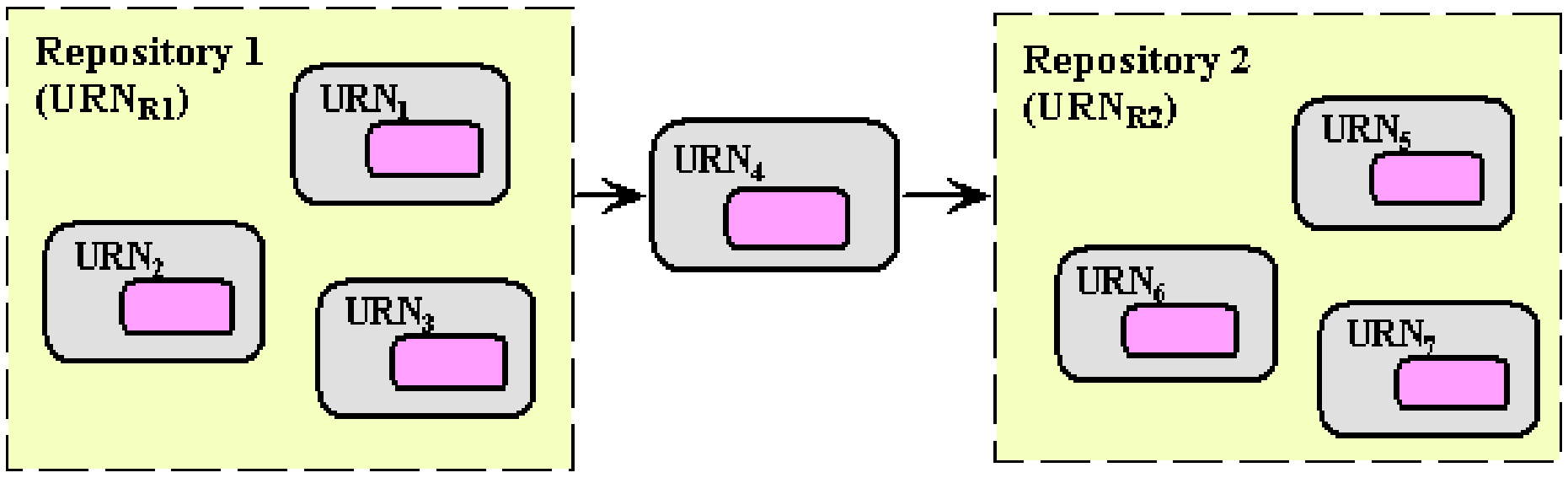

**Fig. 6.** DigitalObject Repositories

*Containment* of DigitalObjects in Repositories is facilitated by a naming service. A DigitalObject effectively does not exist in the infrastructure (i.e., it is not accessible to clients) unless it has a registered identifier such as a URN. Essentially, a DigitalObject is contained in a repository when the name service *resolves* the DigitalObject's identifier to that repository. For example, an object named URN$_1$ is "contained" in a repository URN$_{R1}$, when the name service resolves the name URN$_1$ to the repository named URN$_{R1}$. Once at the repository, client access to the requested DigitalObject is achieved through the `AccessDigitalObject()` request, with URN$_1$ as an argument.

Table 2 shows the set of Repository service requests implemented in FEDORA. We expect to add other utility-oriented operations as we continue to refine the service.

**Table 2.** Repository Service Requests

| For Access: | |
|---|---|
| AccessDigitalObject | Takes a DigitalObject name (e.g. URN) as input, and returns a reference to a DigitalObject. This reference can be used to perform service requests on the DigitalObject via its Primitive Disseminator. |
| **For Management:** | |
| CreateDigitalObject | Creates a new DigitalObject, which is an empty wrapper with a PrimitiveDisseminator. Returns a reference to the DigitalObject that can be used to perform structural manipulations required to populate the empty object. |
| DepositDigitalObject | Takes a reference to a DigitalObject as input, and returns a unique name for that object that can then be registered with the naming service. The DigitalObject name will be registered to resolve the repository that created it. |
| DeleteDigitalObject | Takes a DigitalObject name as input, and removes all references to the DigitalObject from the repository, including its registration in the naming service. |
| ReplicateDigitalObject | Takes a DigitalObject reference and the name of a target repository as input. Creates a replica of the DigitalObject in the target repository by marshalling a byte stream representation of the object from the source repository to the target. Updates the name service to include target repository as a resolve location. |
| MoveDigitalObject | Takes a DigitalObject reference and the name of a target repository as input. Creates a replica of the DigitalObject in the target repository and deletes the object from its source. Updates the name service to reflect new repository location. |

## 7.0 Putting in All Together : *Accessing a Digital Object*

We have introduced the DigitalObject as a two-layered entity, consisting of the DigitalObject structural kernel and the interface layer that exposes its content-type Disseminators. We have introduced the Repository that provides a management layer to the architecture. Once DigitalObjects are created and stored in a Repository, they can be accessed by clients. Figure 7 shows a typical request sequence for interaction with a DigitalObject once it has been accessed in a Repository. The figure assumes that the `AccessDigitalObject(URN1)` request has just been successfully completed on a Repository and the client is interacting with the DigitalObject.

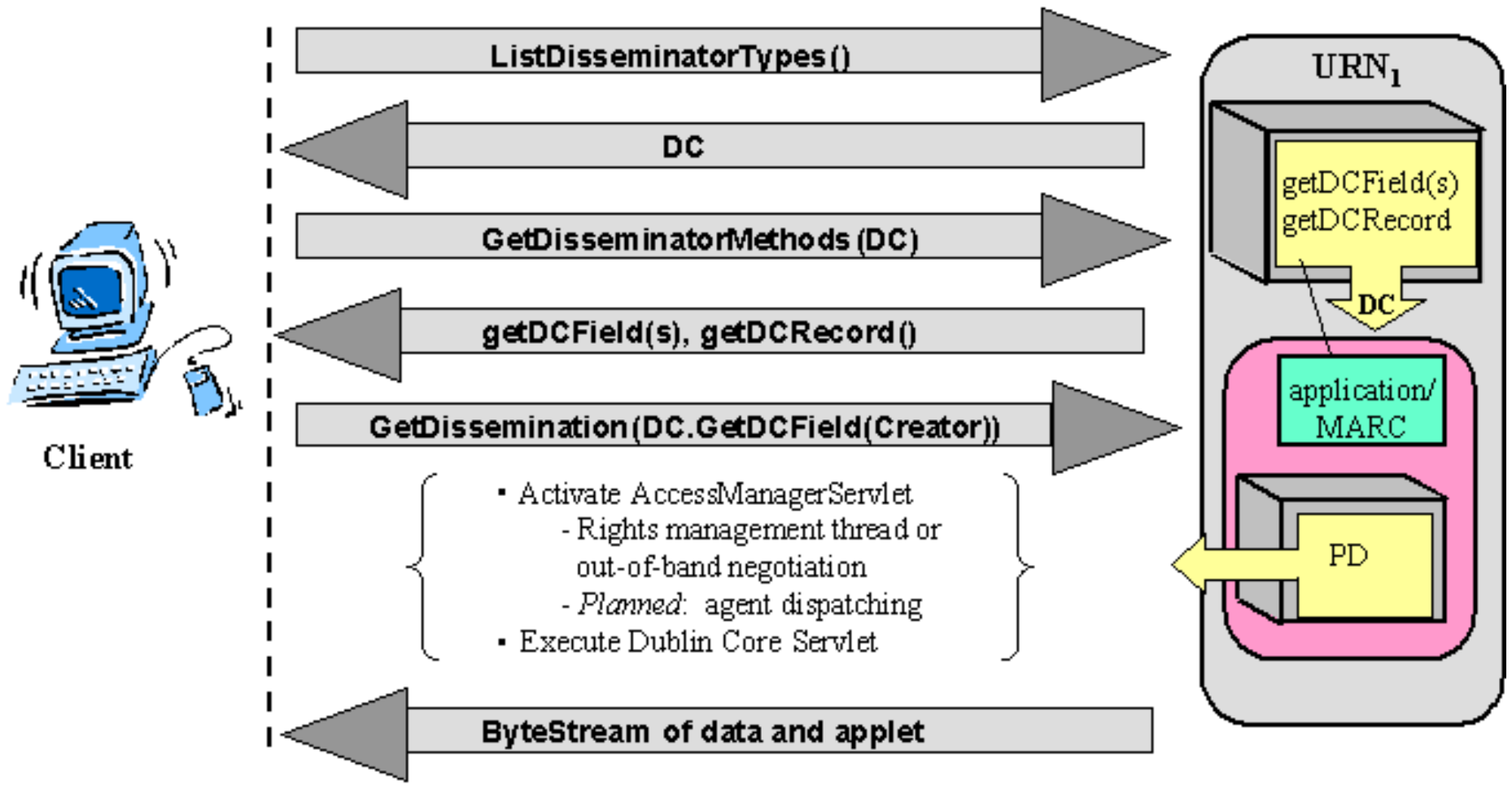

**Fig. 7.** Typical Client Request Sequence after Initial Access of DigitalObject

In the figure, the client is speaking directly to the PrimitiveDisseminator of the DigitalObject. First, the client issues the `ListDisseminatorTypes()` request to find out what content types are available on this DigitalObject. The PrimitiveDisseminator returns only the unique name for the Dublin Core content type, which is `DC`. (If there were multiple Disseminators associated with this object, a list of all the associated content types would be returned.) As in the Web, we expect that clients will be configured to internally handle a set of well-known types. In the case of a well-known type, the client can then immediately invoke a known service request on the DigitalObject; otherwise, the client can ask the object for a list of the service requests for the type. The figure depicts that later scenario, and the client issues the `GetDisseminatorMethods` request with the Dublin Core content-type name as an argument. The request returns a list of method specifications. The client decides to obtain the `Creator` element of the Dublin Core metadata and, therefore, issues a `GetDissemination` request on the object with the appropriate content-type method as an argument. When the request is received by the PrimitiveDisseminator, the AccessManager associated with the `DC` Disseminator is transparently activated, and the content-type service request is processed under its control. (In the figure, AccessManagers are the 3-D boxes surrounding the Disseminators.) The AccessManager may engage the client in an out-of-band negotiation, or it may dispatch an agent to perform an activity (either way it could do things like get username/password clearance, obtain payment, etc.). When the rights management engagement is successful, the AccessManager activates the DisseminatorServlet for the Dublin Core content type. The servlet operates on the MARC DataStream to extract the Dublin Core Creator, and sends this requested data, and a Dublin Core viewing applet back to the client.

## 8.0 Conclusions and Future Work

We have described in this paper a powerful architecture for storing and disseminating digital library content. This architecture makes a number of contributions in the areas of extensibility and interoperability for digital objects and repositories. First, the architecture cleanly separates the structure and raw data stored in a digital object from the semantically meaningful *content types* that are manipulated by clients. It accomplishes this separation through the use of servlets that are able to perform any computable manipulation of the raw data to produce the behaviors defined by the client-visible types. Second, the architecture recursively provides its own type registry, effectively storing types and their mechanisms in DigitalObjects and identifying those types using the unique names (URNs) of those DigitalObjects. Finally, it permits creation and storage of any computable rights management mechanism and the association of these mechanisms with the disseminations defined for a DigitalObject.

At the time of writing this paper, we have completed a prototype implementation of FEDORA that demonstrates the components and services described herein. We will continue to refine the architecture as we test its capabilities in a variety of contexts. In particular, we plan to test how well it accommodates a variety of community-defined content types and rights management mechanisms. Among our test cases will be the interfaces and document model in the current Dienst protocol, the archival types defined by the DLF for the MOA-2 project, and the rights management schemes developed by CNRI in its work for the Library of Congress. As mentioned earlier, we will also conduct a series of interoperability experiments with CNRI.

Since the architecture is distributed by nature, it raises a number of issues related to reliability and security. These are key challenges that we plan to investigate in the next phase of our work with FEDORA. Among the critical areas that we have identified and plan to pursue are mobile code security, rights management policy definitions, fail-safe component design, and reliable replication mechanisms.

**Acknowledgements**

The work described in this paper was funded by the Defense Advanced Research Project Agency under Grant No. MDA 972-96-1-006 with the Corporation for National Research Initiatives. This paper does not necessarily represent the views of CNRI or DARPA. We would like to express our gratitude to William Arms, Christophe Blanchi, and Ed Overly for their thoughtful contributions to this architecture.